\documentclass[a4paper,10pt]{article}
\usepackage[margin=1.05in]{geometry}
\usepackage{multirow}
\usepackage[T1]{fontenc}
\usepackage[utf8]{inputenc}
\usepackage{authblk}
\usepackage{graphicx}
\usepackage{amsmath}
\usepackage{amsfonts}
\usepackage{tikz}
\usepackage{pgfplots}
\usepackage{color}
\usepackage{footnote}
\usepackage[bottom]{footmisc}
\usepackage{csquotes}

\title{HetFHMM: A novel approach to infer tumor heterogeneity using factorial Hidden Markov model}
\author[1]{Gholamreza Haffari\thanks{gholamreza.haffari@monash.edu, corresponding author}}
\author[1]{Zhaoxiang Cai\thanks{zxcai1@student.monash.edu}}
\author[1]{Mohammad S. Rahman\thanks{msrah12@student.monash.edu}}
\author[1]{Ann E. Nicholson\thanks{ann.nicholson@monash.edu}}
\affil[1]{Clayton School of Information Technology, Monash University}

\begin{document}
\maketitle

\begin{abstract}
Cancer arises from successive rounds of mutations which generate tumor cells with different genomic variation i.e. clones. For drug responsiveness and therapeutics, it is necessary to identify the clones in tumor sample accurately. Many methods are developed to infer tumor heterogeneity by either computing cellular prevalence and tumor phylogeny or predicting genotype of mutations. All methods suffer some problems e.g. inaccurate computation of clonal frequencies, discarding clone specific genotypes etc. In the paper, we propose a method, called- HetFHMM to infer tumor heterogeneity by predicting clone specific genotypes and cellular prevalence. To infer clone specific genotype, we consider the presence of multiple mutations at any genomic location. We also tested our model on different simulated data. The results shows that HetFHMM outperforms recent methods which infer tumor heterogeneity. Therefore, HetFHMM is a novel approach in tumor heterogeneity research area.
\end{abstract}

\section{Introduction} \label{Intro}
Cancer is a disease, caused by somatic mutations that accumulate in the genome during the lifetime of human \cite{Iman2014}. Somatic mutations generate tumor cells with different genomic variation within tumor to form carcinoma i.e. cancer. This phenomena is known as \emph{Intra Tumor Heterogeneity} \cite{Ding2012} \cite{Gerlinger2012} \cite{Nik-Zainal2012} \cite{Schuh2012} \cite{Shah2012}. In tumor heterogeneity, each type of cells with distinct genomic structure are known as clones \cite{Ha2013} \cite{Jiao2014}. Caraco et al. \cite{Caraco1998} identifies that drug responsiveness and cancer therapies are clone dependent. To develop more accurate drug and therapy, it is important to predict clonal subpopulations (i.e. clones) within tumor accurately.

Each clone contains distinct genomic variation which is expressed by the precise order of nucleotides of DNA molecules, different from other clones \cite{Ding2013} \cite{Iman2014}. To predict the order of nucleotides of DNA molecules, several techniques are developed. These techniques are known as \emph{DNA sequencing}. \emph{Sanger sequencing} \cite{Sanger1975} is the first approach to predict precise order of nucleotides of DNA molecules. This technique cannot produce complete nucleotide sequence of a DNA molecule. Later a new technique is developed to predict complete nucleotides sequence from DNA molecule called \emph{Next Generation Sequencing} in short \emph{NGS}. NGS produces many short sequences of nucleotides to predict complete nucleotide sequence of a DNA molecule. These short sequences are known as \emph{short reads}. During DNA sequencing, some short reads are deleted or amplified by \emph{E. coli} which is used for cloning DNA fragments. Therefore, sometimes short reads cannot produce complete nucleotide sequence of a DNA molecule. It is a challenging task to infer clones from short reads.

Many researches and methods are developed to infer clones from short reads. These methods detect clones by either estimating cellular prevalence\footnote{The fraction of tumor cell subpopulation represented by a specific clone \cite{Jiao2014}} and tumor phylogeny\footnote{Evolution of clones of a population by mutational processes \cite{Jiao2014}} or inferring genotypes of all mutations. At first researchers develop many methods to infer genotypes of all mutations by considering genotype as the signature of a clone. GPHMM \cite{Li2011} and Apolloh \cite{Ha2013} are the pioneer methods to infer clones from short reads. These methods detect only copy number variation (CNV) in form of genotypes to predict clones by assuming that only one clone exhibits one mutation which appears at a genomic location. Loss of heterogeneity (LOH)\footnote{Loss of heterozygosity (LOH) is a gross chromosomal event that results in loss of the entire gene and the surrounding chromosomal region.} is not predicted together with CNV. TH-HMM \cite{Xia2013} and CLImAT \cite{Yu2014} detect LOH with CNV. But GPHMM \cite{Li2011}, Apolloh \cite{Ha2013}, TH-HMM \cite{Xia2013} and CLImAT \cite{Yu2014} do not consider the presence of multiple clone which do not exhibit a mutation at a location in tumor sample. This important feature is considered in OncoSNP-SEQ \cite{Yau2013} and TITAN \cite{Ha2014} to infer genotypes. Above mentioned methods do not work on presence of multiple mutations at any location.

Roy et al. \cite{Roy2014} predicts the phylogeny relation between clones by using multicolor lineage tracing method alias \emph{Brainbow}. To identify the phylogeny, it is necessary to cluster mutations. TITAN \cite{Ha2014} clusters all mutations into several classes. Other than TITAN \cite{Ha2014}, PyClone \cite{Roth2014}, PhyloSub \cite{Jiao2014}, PhyloWGS \cite{Deshwar2014} and Rec-BTP \cite{Iman2014} identify clusters of mutations by computing and clustering cellular frequencies of all mutations. But these methods cannot identify the clones with same cellular prevalence separately. Like GPHMM \cite{Li2011} and Apolloh \cite{Ha2013}, these methods consider the presence of one mutation at a genomic location.

In real world, many mutations can appear at any genomic location. There are several scenarios appear at any genomic location. (a). Some clones have same mutation which is harboured by their ancestor clone. (b) One clone exhibits a mutation which is identical from others. Genotype of a genomic location cannot reveal types of mutation which appear at that location. Clone specific genotype is a special genotype which express the genotype of a mutation harboured by a clone. Clonal signature\footnote{Expressed by genotypes}, clusters of mutations and types of mutation of a location can be easily predicted by clone specific genotype. No existing method can detect clone specific mutation. This important feature is addressed in our proposed method- \emph{\underline{Het}erogeneity \underline{F}actorial \underline{H}idden \underline{M}arkov \underline{M}odel \emph{(HetFHMM)}}. On the other hand, existing methods detect either cellular prevalence or genotypes. Genotype of a mutation and cellular prevalence combinely affect the count of short reads which are the inputs of existing methods. Our proposed HetFHMM infers clone specific genotypes and cellular prevalences together from the short reads to predict clonal subpopulations within tumor. It is a novel approach in tumor heterogeneity to predict clones within tumor.

Our proposed HetFHMM is discussed in section \ref{model}. We tested our designed method with recent methods for simulated data. It is found that our method outperforms recent methods. We discuss our experiments and results in section \ref{exp}.

\section{HetFHMM} \label{model}
To infer clone specific genotypes and cellular prevalence from short reads, we develop a \emph{factorial hidden Markov model} (FHMM) which is first proposed by Ghahramani et al. \cite{Ghahramani1997}. In FHMM, there are $n$ number of chains. Each chain for individual object. Ghahramani  et al. \cite{Ghahramani1997} first develops FHMM to identify the multiple simultaneous potential speakers from speech signals. In this FHMM, each chain is dedicated for each speaker. In our problem, tumor sample contains multiple clones whose genotypes and cellular prevalence would be different from each other. Like Ghahramani et al. \cite{Ghahramani1997}, we predict clones by $n$ chains using FHMM. In HetFHMM, we consider first chain i.e. chain 0 as normal cells. Rest of other chains represent the clones within tumor sample.

\begin{savenotes}
\begin{table*} \label{tab:geno}
 \caption{Genotype variable space}
 \center
 \scriptsize
 \begin{tabular}{|c|c|c|c|}
  \hline
  copy number & Genotype state & Genotype & Description\\
  \hline
  0 & 0 & $\emptyset$ & Nullizygous\footnote{Both alleles are missing at genomic location.}\\
  \hline
  \multirow{2}{*}{1} & 1 & A & \multirow{2}{*}{hemizygous\footnote{One allele is missing at genomic location.}}\\
  \cline{2-3}
   & 2 & B & \\
  \hline
  \multirow{3}{*}{2} & 3 & AA & Copy neutral with LOH\\
  \cline{2-4}
   & 4 & AB & Normal copy\\
  \cline{2-4}
   & 5 & BB & Copy neutral with LOH\\
  \hline
  \multirow{4}{*}{3} & 6 & AAA & Three copies with LOH\\
  \cline{2-4}
   & 7 & AAB & Three copies with duplication of \textbf{A} allele \\
  \cline{2-4}
   & 8 & ABB & Three copies with duplication of \textbf{B} allele \\
  \cline{2-4}
   & 9 & BBB & Three copies with LOH\\
  \hline
  \multirow{5}{*}{4} & 10 & AAAA & Four copies with LOH\\
  \cline{2-4}
   & 11 & AAAB & Four copies with duplication of \textbf{A} allele \\
  \cline{2-4}
   & 12 & AABB & Four copies with duplication of both alleles \\
  \cline{2-4}
   & 13 & ABBB & Four copies with duplication of \textbf{B} allele \\
   \cline{2-4}
   & 14 & BBBB & Four copies with LOH\\
  \hline
  \multirow{5}{*}{4} & 15 & AAAAA & Five copies with LOH\\
  \cline{2-4}
   & 16 & AAAAB & Five copies with duplication of \textbf{A} allele \\
  \cline{2-4}
   & 17 & AAABB & \multirow{2}{*}{Five copies with duplication of both alleles}\\
  \cline{2-3}
   & 18 & AABBB & \\
  \cline{2-4}
   & 19 & ABBBB & Five copies with duplication of \textbf{B} allele \\
  \cline{2-4}
   & 20 & BBBBB & Five copies with LOH \\
  \hline
 \end{tabular}
\end{table*}
\end{savenotes}

In the model, genotype $G_{t,n}$ of any clone $n$ at $t$th location is one of the hidden variables. Duan et al. \cite{Duan2013} is found in their experiments that copy number varies upto 5. Copy number variation exhibits 21 genotype states are shown in table- \ref{tab:geno}. In the model, we infer another latent variable- cellular prevalence vector $\Phi$ which contains cellular prevalence of all clones of tumor sample.

\begin{figure}
 \centering
 \includegraphics[width=8cm]{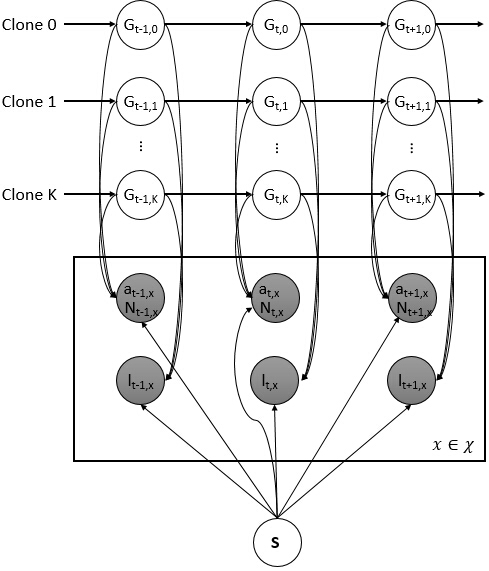}
 \caption{Probabilistic graphical model of HetFHMM}
  \label{fig:PGM}
\end{figure}

Our model takes the number of total short reads $N_t$, reference read counts\footnote{The number of short reads which is matched with reference genome} $a_t$ and log ratio of tumor-normal reads depth $l_t$ at any mutant location $t$ as inputs. Reference read counts $a_t$ follows binomial distribution with total sequence reads $N_t$ and $P_{b_t}(G_t,\Phi)$. The parameter $P_{b_t}(G_t,\Phi)$ computes probability of reference allele, as follows:
\begin{equation}\label{eqn:proBin}
 P_{b_t}(G_t,\Phi) = \frac{\sum_{k=0}^{K}{\phi_{k} . r_{g_{t,k}} . c_{g_{t,k}}}}{\sum_{k=0}^{K}{\phi_{k} . c_{g_{t,k}}}}
\end{equation}
where $K$, $r_{g_{t,k}}$ and $c_{g_{t,k}}$ are the number of clones, reference allele ratio\footnote{$r_{g_{t,k}}=\frac{A_g{_{t,k}}}{c_{g_{t,k}}}$ where $A_{g_{t,k}}$ is the number of A allele of any genotype $G_{t,k}$. For an example, if genotype is AAB, $A_{g_{t,k}}$=2, $c_{g_{t,k}}$=3 and $r_{g_{t,k}}=\frac{2}{3}$} and copy number of clone $k$ at mutant location $t$ respectively. In the model we assume that log ratio of tumor-normal read depth is gaussian distributed with mean $m_t(G_t,\Phi,o)$ and standard deviation $\sigma$. The number of short reads of tumor and normal cells are very large. As per central limit theorem, all large data are normally distributed. For this reason we consider log ratio of tumor-normal read depth is gaussian distributed. Mean of log ratio of tumor-normal read depth is:
\begin{equation} \label{eqn:mean}
 \mu_t(G_t,\Phi,o) = \frac{\sum_{k=0}^{K}{\phi_k . c_{g_{t,k}}}}{\phi_0.c_{g_{t,0}}.\sum_{k=1}^{K}{\phi_k.o}}
\end{equation}
Where $o$ be tumor ploidy parameter which is set to 3. In addition, our proposed model can work on multiple samples $x \in X$ of same patient. Probabilistic graphical model of HetFHMM is shown in fig-\ref{fig:PGM}.

Factorial Hidden Markov Model contains transition and emission probabilities. Like FHMM, transition probability of HetFHMM has been expressed as $P(G_{t,k}=i|G_{t-1,k}=j)\ =\ A_{t,k}(i,j)$. We employ transition probability from the study done by Colella et al.\cite{Colella2007} which captures the position specific effects accurately, is as followed:
\begin{equation} \label{eqn:trans}
 \underbrace{P(G_{t,k}=i|G_{t-1,k}=j)}_{A_{t,k}(i,j)} = \left\{
  \begin{array}{l l}
    \rho_t & \quad \text{if $i$=$j$}\\
    \frac{1-\rho_t}{D_k-1} & \quad \text{Otherwise}
  \end{array} \right.
\end{equation}
where
\begin{equation*}
 \rho_t\ =\ 1-\frac{1}{2}(1-e^{\frac{-d_t}{L}})
\end{equation*}
$L$, $d_t$ and $D_k$ be the average length of the sequence reads\footnote{It was observed to be 2 Megabases (2 $\times\ 10^6$ bases) in 104 breast tumors (rounded to the nearestMb.)}, dimension of state space (in this model it is 21) and distance between the mutant locations $t$ and $t-1$ in clone $k$ respectively. Emission probabilities for generating observation based on hidden variables $G_{t,k}$ and $\Phi$ are as followed:
\begin{equation}
 P(a_t|N_t,G_t,\Phi)\ =\ \mathcal{B}\emph{in}(a_t|N_t,P_{b_t}(G_t,\Phi))
\end{equation}
\begin{equation}
 P(l_t|\sigma,\phi,G_t,\Phi)\ =\ \mathcal{N}(l_t|\mu_t(G_t,\Phi,o),\sigma)
\end{equation}

In order to infer the hidden variables $G$ and $\Phi$, we use negative logarithm of likelihood function to pursue $G$ and $\Phi$ which give the highest joint probability of the model.
 \begin{align}
 P(G,l,a,\Phi|N,\o) = & \prod_{k=0}^{K}{\prod_{t=0}^{T}{A_t(G_{t,k}|G_{t-1,k})}} \nonumber \\
 & \prod_{x \in X}{\prod_{t=0}^{T}{\mathcal{B}\emph{in}(a_{t,x}|N_{t,x},P_{b_{t,x}}(G_t,\Phi))\mathcal{N}(l_{t,x}|\mu_{t,x}(G_t,\Phi,o),\sigma)}}
 \end{align}
Since the exact inference for FHMM is intractable \cite{Ghahramani1997}, we use a sampling method to get an approximation. Markov chain Monte Carlo (MCMC) sampling is widely adopted for this task. Gibbs sample in one of the simple sampling schemes among MCMC methods. In order to run Gibbs sampling, we need to start with an initial model in which all genotypes are randomly generated. Except for the normal chain, we randomly choose a genotype for each variable based on the uniform distribution. Then, each hidden variable is sampled given the current state of rest of the variables. In our case, the probability of each genotype for a hidden variable $G_{t,k}$ is :
\begin{equation}
 P(G_{t,k}) \propto A_t(G_{t,k}|G_{t-1,k})A_{t+1}(G_{t,k}|G_{t-1,k})\mathcal{B}\emph{in}(a_{t,x}|N_{t,x},P_{b_{t,x}})\mathcal{N}(l_{t,x}|\mu_{t,x},\sigma)
\end{equation}
We sample each current state given another four variables: the previous genotype $G_{t-1}$, the next genotype $G_{t+1}$ and two observations $l_t$ and $a_t$.

Having fixed all the values for genotype variables $G$, the negative log likelihood can be minimized using Exponentiated Gradient Descent, which is similar to normal Gradient Descent. Gradient Descent is an algorithm to find the minimum of a given objective function (target function) by gradually approaching the minimum along the direction of the gradient function. Exponentiated Gradient (EG) \cite{Kivinen1997} algorithm is a variant of normal Gradient Descent. The difference is that the update for Gradient Descent is to subtract the gradient of a target function, where as in EG the update is done by multiplying the exponents of the negative gradient. So in EG we pursue:
\begin{equation}
 \max_{\Phi \in \triangle} - \mathcal{L}(\Phi)
\end{equation}
that is $\sum_{k}{\phi_k}=1$ and $\phi_k \ge 0$ where $\mathcal{L}$ denotes the objective function. In addition, it is proved to perform better when the target is sparse. In other words, it allows us to identify clones even if it contains only a very small proportion of cells. In our case, the objective function is the log likelihood function of the model, i.e. $\mathcal{L}(\Phi)\ =\ \log(P(G,l,a,\Phi|N,o))$. To solve the above maximization problem, the EG updates are as follows:
\begin{equation}
 \phi_k^{new} = \phi_k e^{-(\eta \nabla_{\phi_k}\mathcal{L}(\Phi))}
\end{equation}
where $\eta$ is the learning rate. After updating each component of the latent vector $\Phi$, the values are normalized so that they sum to one. In our model, for the EG updates, we need the derivatives which are derived using the chain rule, as follows:
\begin{align}
 \mathcal{L}(\Phi)= & \sum_{t}\log {N_t \choose a_t}+a_t\log P_{b_t}+(N_t-a_t)\log(1-P_{b_t}) \nonumber\\
 & + \log(\frac{1}{\sigma\sqrt{2\pi}})-\frac{(l_t-\mu_t)^2}{2\sigma^2}+\emph{const}
\end{align}
\begin{equation}
 \nabla \mathcal{L}(\Phi)=\sum_{t}{[(\frac{a_t}{P_{b_t}}-\frac{N_t-a_t}{1-P_{b_t}})\cdot \nabla\mu_t+\frac{l_t-\mu_t}{\sigma^2}\cdot \mu_t]}
\end{equation}
\begin{equation}
 \frac{dP_{b_t}}{d\phi_k}=\frac{c_{t,k}\cdot (r_{t,k}-P_{b_t})}{\sum_{k=0}^{K}{\phi_k\cdot c_{t,k}}}
\end{equation}
\begin{equation}
 \frac{d\mu_t}{d\phi_0}=\frac{c_{t,0}\cdot (1-\mu_t)}{\phi_0\cdot c_{t,0}+\sum_{k}{\phi_k\cdot o}}
\end{equation}
\begin{equation}
 \frac{d\mu_t}{d\phi_k}=\frac{c_{t,0}\cdot o \mu_t}{\phi_0\cdot c_{t,0}+\sum_{k}{\phi_k\cdot \phi}}
\end{equation}
Substitute $\frac{dP_{b_t}}{d\phi_k}$ and $\frac{d\mu_t}{d\phi_k}$ back to $\mathcal{L}(\Phi)$, we can get the gradient of the objective function with respect to variable $\Phi$.

It is known that Gibbs sampling stops when the convergence criteria is met. In our case, we define the number of times that each genotype and cellular prevalence of $n$ clones are sampled as the convergence criteria.

\section{Experiments and results} \label{exp}
We carry out several experiments to HetFHMM on simulated data to evaluate its inference of clone specific genotypes at any mutant location including cellular prevalence. We generate simulated data from factorial Hidden Markov model with the number of clones varies from 3, 4, 5 and 6. The cellular prevalence vector $\Phi$ is specified before generating FHMM. For each FHMM, we select the number of mutant location by random which varies from 1k to 10k. After selecting the location, first we proceed to chain 0 which represents as normal cells with genotype \textbf{AB}. We assume that hidden variables for all chains at first location are also \textbf{AB}. Next, the genotypes of each clone at all mutant locations are generated using equation-\ref{eqn:trans}. We set the expected probability of reference allele $P_b$ and the mean $\mu_t$ of $l_t$ by using equations-\ref{eqn:mean} and \ref{eqn:proBin} respectively. Finally we generate observation variables: reference allele read counts and log ratio of tumor-normal contents by random production of total read counts which varies from 1k~10k.

We implement our model HetFHMM on these simulated data. For comparing our model with recent methods, we also implement PyClone \cite{Roth2014}, PhyloSub \cite{Jiao2014} and Rec-BTP \cite{Iman2014} on these data. HetFHMM gives two outputs: cellular prevalence and clone specific genotypes. Whereas PyClone \cite{Roth2014}, PhyloSub \cite{Jiao2014} and Rec-BTP \cite{Iman2014} predict cellular prevalence and clusters of all mutations. No method infers clone specific genotypes from the short reads. Clusters of all mutations can be determined from clone specific genotypes.  Similar types of output of HetFHMM are also generated by PyClone \cite{Roth2014}, PhyloSub \cite{Jiao2014} and Rec-BTP \cite{Iman2014}. For this reason we compare the results of HetFHMM with PyClone \cite{Roth2014}, PhyloSub \cite{Jiao2014} and Rec-BTP \cite{Iman2014}.

We use $V\mbox{-}measure$ to assess the clusters of all mutation that are either generated from clone specific genotypes or PyClone \cite{Roth2014}, PhyloSub \cite{Jiao2014} and Rec-BTP \cite{Iman2014} with gold standard. $V\mbox{-}measure$ is an entropy-based external cluster validation measure to evaluate predicted clusters with respect to gold standard classes. It measures how successfully the criteria of homogeneity\footnote{Each cluster contains only members of a single class \cite{Rosenberg2007}.} and completeness\footnote{All members of a given class are assigned to the same cluster \cite{Rosenberg2007}.} have been satisfied. $V\mbox{-}measure$ is the only external entropy based measure which successfully and efficiently evaluate cluster prediction than any other cluster evaluation measures. It is also used in PyClone to evaluate its clustering methodology. Rosenberg et al. \cite{Rosenberg2007} defined homogeneity as followed:
\begin{equation}
 h = 1-\frac{H(N|n)}{H(N)}
\end{equation}
where $H(N|n)$ is the conditional entropy of the class $N$ given the cluster $n$, as followed:
\begin{equation*}
 H(N|n)=-\sum_{k=1}^{K}{\sum_{c=1}^{C}{\frac{a_{ck}}{M}\log{\frac{a_{ck}}{\sum_{c=1}^{C}{a_{ck}}}}}}
\end{equation*}
and $H(N)$ is the entropy of the class $N$, as followed:
\begin{equation*}
 H(N)=-\sum_{c=1}^{C}{\frac{\sum_{k=1}^{K}{a_{ck}}}{M}\log{\frac{\sum_{k=1}^{K}{a_{ck}}}{M}}}
\end{equation*}

$C$, $K$, $M$, $N$, and $n$ and $a_{c,k}$ are the number of gold standard clones, the number of predicted clones, the number of mutations, a gold standard clone, a predicted clone  and the number of mutations that are members of gold standard clones $c$ and predicted clones $k$. They also defined the completeness as followed:
\begin{equation}
   c = 1-\frac{H(n|N)}{H(n)}
\end{equation}
where $H(n|N)$ is the conditional entropy of the cluster $n$ given the class $N$, as followed:
\begin{equation*}
   H(n|N)=-\sum_{c=1}^{C}{\sum_{k=1}^{K}{\frac{a_{ck}}{M}\log{\frac{a_{ck}}{\sum_{k=1}^{K}{a_{ck}}}}}}
\end{equation*}
and $H(n)$ is the entropy of the cluster $N$, as followed:
\begin{equation*}
   H(n)=-\sum_{k=1}^{K}{\frac{\sum_{c=1}^{C}{a_{ck}}}{M}\log{\frac{\sum_{c=1}^{C}{a_{ck}}}{M}}}
\end{equation*}

$V\mbox{-}measure$ is defined as the harmonic mean of distinct homogeneity and completeness scores by Rosenberg et al. \cite{Rosenberg2007}:
\begin{equation}
  V\mbox{-}measure=\frac{2\times h \times c}{h + c}
 \end{equation}

The value of $V\mbox{-}measure$ ranges from 0 to 1. If its value is 1, it means perfect matching; otherwise not. For assessing our inferred clone specific genotype with the clusters of mutations by $V\mbox{-}measure$, we generate a two dimensional matrix $M_{K\times T}^{F}$ in which if any mutation $t$ is present in any clone $k$ means the genotype of $t$ mutation in clone $k$ is other than \textbf{AB}, $M_{k,t}^{F}=1$; Otherwise zero. We also produce similar matrices $M_{K\times T}^{G}$, $M_{K\times T}^{RT}$, $M_{K\times T}^{PC}$ and $M_{K\times T}^{PS}$ for gold standard, Rec-BTP \cite{Iman2014}, PyClone \cite{Roth2014} and PhyloSub \cite{Jiao2014} respectively. Generated genotypes that are used to produce simulated data, are considered as gold standard.

\begin{figure}[htbp]
\center
\begin{tikzpicture}
\begin{axis}[
width=0.75\linewidth,
height=0.25\linewidth,
ybar,
ylabel={$V\mbox{-}measure$},
bar width=0.25cm,
symbolic x coords={3 Clones,4 Clones,5 Clones,6 Clones},
xtick=data,
legend style={
    at={(0.5,-0.3)},
    anchor=north,legend columns=-1
},
ymin=0,
ytick={0.0,0.25,0.5,0.75,1.0},
]
\addplot[style={red,fill=red,mark=none}] coordinates {(3 Clones,0.7415) (4 Clones,0.4391) (5 Clones,0.4103) (6 Clones,0.4211)};
\addplot[style={yellow,fill=yellow,mark=none}] coordinates {(3 Clones,0.5210) (4 Clones,0.3412) (5 Clones,0.3387) (6 Clones,0.3465)};
\addplot[style={green,fill=green,mark=none}] coordinates {(3 Clones,0.2989) (4 Clones,0.2769) (5 Clones,0.2865) (6 Clones,0.2753)};
\addplot[style={blue,fill=blue,mark=none}] coordinates {(3 Clones,0.1566) (4 Clones,0.1450) (5 Clones,0.1405) (6 Clones,0.1422)};
\legend{HetFHMM,PhyloSub,PyClone,Rec-BTP}
\end{axis}
\end{tikzpicture}
\caption{$V\mbox{-}measure$ result comparison among \emph{HetFHMM}, \emph{Rec-BTP}, \emph{PyClone} and \emph{PhyloSub} algorithms for 3, 4, 5 and 6 clones synthetic dataset.}
\label{fig:mutClusSyn}
\end{figure}
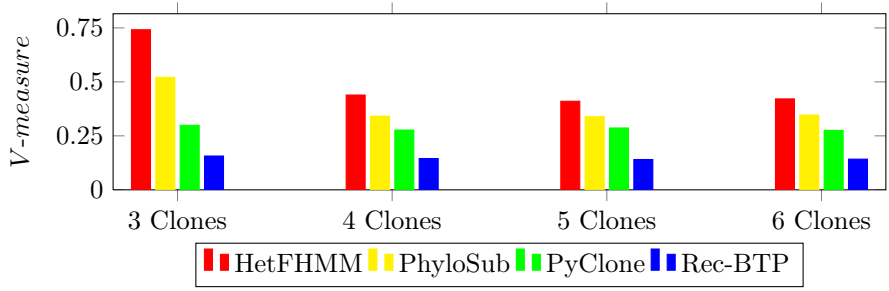

Using these matrices, we compute $V\mbox{-}measure$ of the clustered outputs of HetFHMM, PyClone \cite{Roth2014}, PhyloSub \cite{Jiao2014} and Rec-BTP \cite{Iman2014} with respect to gold standard. The cluster validation result is shown in fig.-\ref{fig:mutClusSyn}. It clearly shows that HetFHMM outperforms Rec-BTP \cite{Iman2014}, PyClone \cite{Roth2014} and PhyloSub \cite{Jiao2014}. PyClone \cite{Roth2014}, PhyloSub \cite{Jiao2014} and Rec-BTP \cite{Iman2014} produce more clones than original number of clones which affects the completeness and homogeneity of $V\mbox{-}measure$. More over, from fig.-\ref{fig:mutClusSyn}, it is also found that increasing number of clones inversely affects the $V\mbox{-}measure$ of each method.

We evaluate the clonal frequencies by using $RMSD$ value after examining the clusters of mutation. \emph{Root mean square distance} or $RMSD$ is used to compute the distance/ gap between cellular prevalence of predicted clones and gold standard clones. Many predicted clones contains some of mutations of a gold standard clone. Among them, a predicted clone would contain maximal number of mutations of gold standard clones. This predicted clone is called significant clone of a gold standard clone. This is why, clonal frequencies of all predicted clones are not compared with gold standard. Before computing $RMSD$ value, we obtain significant clones from predicted clones. For detecting significant clones, we set a value called threshold. If $V\mbox{-}measure$ of any predicted clone $n$ and gold standard clone $N$ is above threshold, the predicted clone $n$ is considered as significant clone of gold standard clone $N$. $RMSD$ value is calculated by:
\begin{equation}
 \begin{array}{l l}
 RMSD\ = & \sqrt{\frac{1}{S}\sum_{\forall significant\ clones}{\|\Phi_n-\Phi_N\|^2}}\\
  & \text{if $n$ is significant clone of $N$, detected by $V\mbox{-}measure$}
 \end{array}
\end{equation}
$S$ be the number of significant predicted clones. The zero value of $RMSD$ indicates the perfect computation of clonal frequencies. But, most of the cases, $RMSD$ value of different methods changes significantly. For this reason, we use $\log{RMSD}$ instead of $RMSD$. $\log{RMSD}$ value which helps us to understand the cellular prevalence comparing results better than $RMSD$.

\begin{figure}[htbp]
\center
\begin{tikzpicture}
\begin{axis}[
width=0.75\linewidth,
height=0.35\linewidth,
ybar,
ylabel={$\log{RMSD}$},
bar width=0.25cm,
symbolic x coords={3 Clones,4 Clones,5 Clones,6 Clones},
xtick=data,
legend style={
    at={(0.5,-0.3)},
    anchor=north,legend columns=-1
},
ymin=0,
ytick={0.0,0.5,1.0,1.5,2.0,2.5,3.0},
]
\addplot[style={red,fill=red,mark=none}] coordinates {(3 Clones,0.1977) (4 Clones,0.6906) (5 Clones,0.6582) (6 Clones,0.7565)};
\addplot[style={yellow,fill=yellow,mark=none}] coordinates {(3 Clones,0.6228) (4 Clones,0.8312) (5 Clones,0.9621) (6 Clones,1.3980)};
\addplot[style={green,fill=green,mark=none}] coordinates {(3 Clones,2.1587)(4 Clones,2.6712) (5 Clones,2.6169) (6 Clones,2.3782)};
\addplot[style={blue,fill=blue,mark=none}] coordinates {(3 Clones,2.0451) (4 Clones,2.2716) (5 Clones,2.6300) (6 Clones,2.4791)};
\legend{HetFHMM,PhyloSub,PyClone,Rec-BTP}
\end{axis}
\end{tikzpicture}
\caption{$\log{RMSD}$ comparison among \emph{HetFHMM}, \emph{PhyloSub}, \emph{PyClone} and \emph{Rec-BTP} algorithms for 3, 4, 5 and 6 clones synthetic dataset.}
\label{fig:RMSDSyn}
\end{figure}
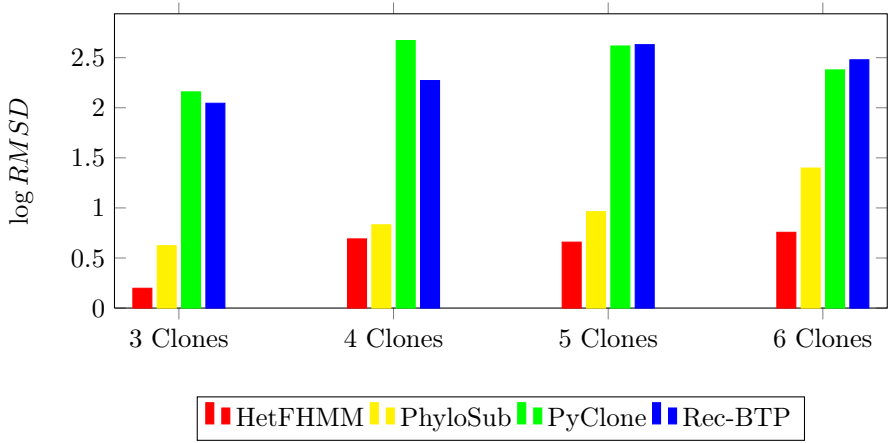
In cellular prevalence evaluation, we set the value of threshold to 0.25. The comparative result of $\log{RMSD}$ of HetFHMM, Rec-BTP \cite{Iman2014}, PyClone \cite{Roth2014} and PhyloSub \cite{Jiao2014} are shown in fig.-\ref{fig:RMSDSyn}. We find that HetFHMM produces less $\log{RMSD}$ value compare to Rec-BTP \cite{Iman2014}, PyClone \cite{Roth2014} and PhyloSub \cite{Jiao2014}. The $V\mbox{-}measure$ comparison shows that mutation clustering by Rec-BTP \cite{Iman2014}, PyClone \cite{Roth2014} and PhyloSub \cite{Jiao2014} are not more accurate than HetFHMM. It affects the number of mutations in a clone which affects the cellular prevalence of recent methods.

\section{Conclusion}
HetFHMM can detect clones by genotypes and cellular prevalence together. We develop HetFHMM by considering the presence of multiple clones and mutations at any genomic location which helps to find the more accurate clones and their cellular prevalence than existing methods: PyClone \cite{Roth2014}, PhyloSub \cite{Jiao2014} and Rec-BTP \cite{Iman2014}. Therefore, it can be said that HetFHMM is a novel approach in tumor heterogeneity research to infer clones.

Despite improved performance of HetFHMM, it has some issues. We consider any mutation is dependant on previous mutation. In the real world, all previous mutations have position specific affects on one mutation which is not considered in this model. The number of clones within a tumor sample is predefined, but it needs to be predicted from tumor sample. On the other hand, we experiment our model only on simulated data. We have a plan in future to test our model on real large tumor sample. We also improve our HetFHMM by considering above mentioned important features.

\end{document}